\documentclass[twocolumn,showpacs,prd]{revtex4}

\usepackage{dcolumn}
\usepackage{bm}
\usepackage{graphicx}

\begin{document}

\title{Electric Dipole Moments and Polarizability \\ 
in the Quark-Diquark Model of the Neutron}

\author{Y.N. Srivastava$^{(a,b,c)}$, A. Widom$^{(b)}$, J. Swain$^{(b)}$ and O. Panella$^{(c)}$}
\affiliation{$^{(a)}$Physics Department, University of Perugia, Via A. Pascoli, 06123 Perugia, Italy} 
\affiliation{$^{(b)}$Physics Department, Northeastern University, 110 Forsyth St. Boston, MA 02115, USA}
\affiliation{$^{(c)}$INFN, Sezione di Perugia, Via A. Pascoli, 06123 Perugia IT}

\begin{abstract}
For a bound state internal wave function respecting parity symmetry, 
it can be rigorously argued that the mean electric dipole moment must 
be strictly zero. Thus, both the neutron, viewed as a bound state of three 
quarks, and the water molecule, viewed as a bound state of ten 
electrons two protons and an oxygen nucleus, both have zero mean electric dipole 
moments. Yet, the water molecule is said to have a {\em nonzero} dipole moment 
strength $d=e\Lambda $ with $\Lambda_{H_2O} \approx 0.385\ \dot{A}$. The neutron 
may also be said to have an electric dipole moment strength with 
$\Lambda_{neutron} \approx 0.612\ fm$. The neutron analysis can be made 
experimentally consistent, if one employs a quark-diquark model of neutron 
structure.   
\end{abstract}


\maketitle

\section{Introduction \label{intro}}

Consider an internal ground state wave function 
\begin{math} \Psi \end{math} bound by Hamiltonian 
\begin{math} {\cal H} \end{math} which 
respects parity \begin{math} {\cal P} \end{math} symmetry, 
i.e. 
\begin{equation}
\left[{\cal H},{\cal P}\right]=0.
\label{intro1}
\end{equation}  
The ground state is expected to be in a parity eigenstate 
\begin{equation}
{\cal P}\Psi =\pm \Psi 
\label{intro2}
\end{equation}  
yielding a {\em null} ground state mean electric dipole moment 
\begin{eqnarray}
\overline{\bf d}=\left(\Psi , {\bf d} \Psi \right)
=\left({\cal P}\Psi , {\bf d} {\cal P}\Psi \right),
\nonumber \\ 
\overline{\bf d}
=\left(\Psi , {\cal P}{\bf d} {\cal P}\Psi \right)
=-\left(\Psi , {\bf d} \Psi \right),
\nonumber \\ 
\overline{\bf d}=-\overline{\bf d}=0. 
\label{intro3}
\end{eqnarray}  

\begin{figure}[tp]
\begin{centering}
\includegraphics[scale=0.5]{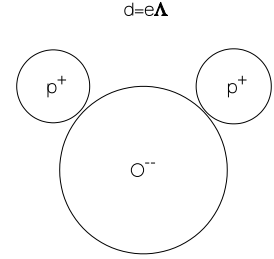}
\caption{Shown is a schematic picture of a water molecule with two protons each of 
charge $+e$ skating on the spherical electronic surface of an xygen ion of 
charge $-2e$. The instantaneous dipole moment strength is 
$d=e\Lambda=2e\tilde{L}\cos(\theta /2)$ wherein $\theta $ is the angle 
between the hydrogen bonds and $\tilde{L}$ is the bond length. The vanishing of 
the mean vector electric dipole moment $\overline{\bf d}=0$ is due to the averaged 
tumbling rotations of the molecule as a whole.}
\label{fig1}
\end{centering}
\end{figure}

The vanishing mean electric dipole moment holds true for both a 
water molecule \begin{math} H_2O  \end{math} and a neutron 
\begin{math} n=udd  \end{math}. The water molecule is a bound state 
of charged constituents, i.e. ten electrons \begin{math} e^- \end{math}, 
two protons \begin{math} p^+=\ ^1 _1H  \end{math} and an oxygen nucleus    
\begin{math} ^{16} _{\ 8}O  \end{math}. The neutron too is a bound state 
of charged constituents, i.e. one up {\it u} and two down {\it d} quarks. 
For both of these bound states, the respect of parity symmetry 
requires a vanishing mean dipole moment 
\begin{math} \overline{\bf d}=0 \end{math}. The search for 
a possible non-vanishing \begin{math} \overline{\bf d} \end{math} for the 
neutron has turned into a high precision experimental sport\cite{Ramsey:1982}, 
yielding  
\begin{math} |\overline{\bf d}_{neutron}/e|<3\times 10^{-13}\ fm \end{math}.
Albeit the vanishing of the mean dipole moment, it is possible to define 
a dipole strength \begin{math} d \end{math} by 
\begin{equation}
d^2=\overline{\bf d\cdot d}=e^2\Lambda^2.
\label{intro4}
\end{equation}

For the case of a water molecule, one attributes an electric 
dipole moment strength  wherein\cite{Clough:1973}  
\begin{equation}
d_{H_2O}/e = \Lambda_{H_2O}\approx 3.85\times 10^{-9}\ cm.
\label{intro5}
\end{equation}
One can understand the notion of an electric dipole strength of a water 
molecule by examining a {\em snap shot} as shown in FIG. \ref{fig1}.  

In Sec.\ref{rotw} the experimental reality of FIG. \ref{fig1} will be explored 
in terms of the rotational energy spectra of the tumbling water molecule. The 
energy-angular momentum levels of the water molecule form non-relativistic 
rotational Regge trajectories\cite{Alpharo:1965} from which one may deduce the 
molecular tensor moment of inertia \begin{math} {\sf I}  \end{math}. The 
eigenvalues of the moment of inertia tensor are consistent with the hydrogen bonding 
picture as can be viewed in FIG. \ref{fig1}.   

\begin{figure}[ht]
\begin{centering}
\includegraphics[scale=0.6]{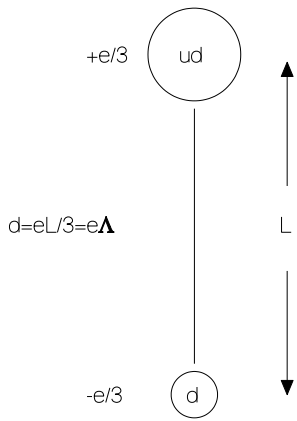}
\caption{Shown is a model of a neutron consisting of a bound state of 
three quarks $n=udd$. Two of the quarks are bound into a diquark ($ud$) 
at one end of a string which in turn is connected to the third ($d$) quark. 
The diquark has charge $e/3$ while the single quark has charge $-e/3$ 
yielding a dipole moment strength of $d=e\Lambda$ with $\Lambda=L/3$. 
The vanishing of the mean vector electric dipole moment $\overline{\bf d}=0$ 
is due to the averaged tumbling rotations of the string as a whole.}
\label{fig2}
\end{centering}
\end{figure}

For the case of a neutron, we shall show in what follows that 
one may attribute to the neutron an electric 
dipole moment strength,  
\begin{equation}
d=e\Lambda , 
\label{intro6}
\end{equation}
wherein   
\begin{equation}
\Lambda_{neutron}\approx 6.120\times 10^{-14}\ cm.
\label{introt}
\end{equation}
One can understand the notion of an electric dipole strength of a neutron  
by examining a {\em snap shot} as shown in FIG. \ref{fig2}.   

The model of a neutron here of interest\cite{Srivastava:2000,Anselmino:1993} 
in what follows consists of a \begin{math} ud  \end{math} diquark at one end 
of a string of length \begin{math} L \end{math} and tension 
\begin{math} \tau \end{math} and a \begin{math} u \end{math} quark at the 
other end of the string. The mean electric dipole moment strength is thereby 
\begin{equation}
d=e\Lambda_{neutron}\ \ \ \Rightarrow 
\ \ \ \Lambda_{neutron}=\frac{L}{3}.
\label{intro7}
\end{equation}
The vanishing of the mean vector neutron electric dipole moment 
\begin{math} \overline{\bf d}=0 \end{math} is due to the quantum tumbling 
rotational motion of the neutron as a whole.

In Sec.\ref{rotn} the experimental reality of FIG. \ref{fig2} will be explored 
in terms of the rotational energy spectra of the tumbling neutron. The 
energy-angular momentum levels of the \begin{math} udd \end{math} system 
form a relativistic rotational Regge trajectory  with a linear mass squared 
versus angular momentum\cite{Gervais:1980,Martin:1970}.

For both the water molecule and the neutron, the experimental value of the 
electric dipole moment strength \begin{math} d=e\Lambda  \end{math} may be 
found from measurements of the polarizability \begin{math} \alpha \end{math}. 
Experimental and theoretical values of \begin{math} \alpha \end{math} are discussed 
in the concluding Sec. \ref{conc}, wherein strong evidence is presented in favor 
of the quark-diquark view as shown in FIG. \ref{fig2}. The diquark should thereby 
be a central notion when classifying hadrons employing quark models.  

\section{Rotational Bands of Water\label{rotw}}

The polarizability \begin{math} \alpha_T  \end{math} at temperature 
\begin{math} T  \end{math} of a classical electric dipole moment 
\begin{math} {\bf d} \end{math} fixed in magnitude and varying in direction 
may be found from the classical fluctuation response theorem for 
Eq.(\ref{intro4}); It is\cite{Debye:1929} 
\begin{equation}
\alpha_T({\rm classical})=\frac{d^2}{3k_BT}\ .
\label{rotw1}
\end{equation}
Eq.(\ref{rotw1}) has yielded an experimental method of defining the dipole moment 
strength in terms of the temperature variations of the polarizability
\begin{equation}
d_{H_2O}^2=-3k_BT^2\frac{d\alpha_T}{dT}=e^2\Lambda_{H_2O}^2
\label{rotw2}
\end{equation}
in a regime wherein \begin{math} d  \end{math} itself does not depend on 
temperature. The experimental number in Eq.(\ref{intro5}) employed  
Eq.(\ref{rotw2}) in the analysis of the data. In and by itself, this analysis 
does not lead to the classic picture of a water molecule pictured in 
FIG. \ref{fig1}. To understand the physical picture of the molecule one must 
consider the rotational energy spectrum of a water molecule which amounts to a 
nonrelativistic  Regge trajectory.

For a total angular momentum \begin{math} {\bf J} \end{math}, 
\begin{equation}
{\bf J\cdot J}=\hbar^2 j(j+1),
\label{rotw3}
\end{equation}
the tumbling water molecule rotational energy levels have the rigid body form 
\begin{equation}
{\cal H}_{\rm rotation}={\bf J}\cdot ({2\ \sf I})^{-1}\cdot {\bf J}=
\frac{J_1^2}{2I_1}+\frac{J_2^2}{2I_2}+\frac{J_3^2}{2I_3}\ ,
\label{rotw4}
\end{equation}
wherein \begin{math} {\sf I}  \end{math} is the moment of inertia tensor with 
principal values \begin{math} I_1<I_2<I_3 \end{math}. The rotational energy 
eigenvalues 
\begin{equation}
{\cal H}_{\rm rotation}\left|j,n\right>={\cal E}_{jn}\left|j,n\right>, 
\label{rotw5}
\end{equation}
give rise to measurable molecular tumbling frequencies,  
\begin{equation}
\omega_{jn.j^\prime n^\prime}=\frac{1}{\hbar }
\left({\cal E}_{jn}-{\cal E}_{j^\prime n^\prime}\right), 
\label{rotw7}
\end{equation}
from which the principal values \begin{math} I_1<I_2<I_3 \end{math} can be deduced. 
In this manner the picture in FIG. \ref{fig1} can be completed with numbers for the 
hydrogen bonding lengths and angles consistent with the numbers for the dipole 
strength.
 
\section{Relativistic Regge Trajectories\label{rotn}}

The neutron analysis proceeds in close analogy to the water molecule 
case. One must consider the rotational states of the \begin{math} udd \end{math} 
quark system which in the zero quark mass limit resides in the string 
of FIG. \ref{fig2}. The tumbling string frequencies obey 
\begin{equation}
\omega =\frac{d \cal E}{dJ}=\frac{\pi c}{L}.
\label{rotn1}
\end{equation}
The string tension is the energy per unit string length, 
\begin{equation}
\tau =\frac{\cal E}{L}\ ,
\label{rotn2}
\end{equation}
so that 
\begin{equation}
\frac{1}{\cal E}\left(\frac{dJ}{d \cal E}\right)=\frac{1}{\pi c \tau }.
\label{rotn3}
\end{equation}
The solution of Eq.(\ref{rotn3}) 
\begin{equation}
J=J_0+\frac{{\cal E} ^2}{2\pi c \tau}\ .
\label{rotn4}
\end{equation}
is thereby the linear relativistic Regge trajectory for the string rotational energy 
levels for the \begin{math} udd \end{math} system.

The ground rotational state of the \begin{math} udd  \end{math}  system is the neutron with 
angular momentum \begin{math} (\hbar/2) \end{math} and mass \begin{math} M \end{math}. The 
first excited rotational state of the \begin{math} udd  \end{math}  system is the delta with 
angular momentum \begin{math} (3\hbar/2) \end{math} and mass \begin{math} M_\Delta \end{math}.  
The experimental masses of the neutron and the delta determine the string tension 
\begin{math} \tau \end{math} employing Eq.(\ref{rotn3}). It is  
\begin{equation}
\frac{2\pi \hbar  \tau}{c^3} =M_\Delta ^2-M^2 =\frac{1}{\alpha^\prime }\ ,
\label{rotn7}
\end{equation}
wherein \begin{math} \alpha^\prime  \end{math} is the conventional Regge slope 
parameter. On the other hand, the length \begin{math} L \end{math} of the neutron 
string in the quark-diquark model follows from Eq.(\ref{rotn2}) to be determined by  
\begin{math} \tau \end{math} via Eq.(\ref{rotn4}). It is  
\begin{equation}
L=\frac{Mc^2}{\tau}=2\pi \left(\frac{\hbar }{Mc}\right)
\left(\frac{M^2}{M_\Delta ^2-M^2}\right)=3\Lambda.
\label{rotn8}
\end{equation}
Thus, the dipole strength of the neutron in the quark-diquark model of 
FIG. \ref{fig2} is given by 
\begin{math} \tau \end{math} via Eq.(\ref{rotn4}); It is   
\begin{eqnarray}
\frac{d_{neutron}}{e}=\Lambda_{neutron}
\nonumber \\ 
\Lambda_{neutron}=\frac{2\pi }{3}
\left(\frac{\hbar }{Mc}\right)
\left[\frac{1}{(M_\Delta /M)^2-1}\right].
\label{rotn9}
\end{eqnarray}
Employing the experimental numbers 
\begin{eqnarray}
\frac{\hbar }{Mc}=0.21009416\ fm\ \ \ {\rm and}
\ \ \ \frac{M_\Delta }{M}=1.311,
\nonumber \\ 
{\rm yielding}\ \ \ \Lambda_{neutron}
\approx 0.6120\ fm, 
\label{rotn10}
\end{eqnarray} 
one finds the neutron dipole strength as in Eq.(\ref{introt}).

Now let us consider the neutron polarizability \begin{math} \alpha  \end{math}. The 
quantum mechanical expression for \begin{math} \alpha  \end{math} is 
\begin{equation}
\alpha =\frac{2}{3}\sum_n\frac{
\left|\left<n \right|{\bf d}\left|0\right> \right|^2}
{{\cal E}_n-{\cal E}_0}\ .
\label{polar}
\end{equation}
For the problem at hand, \begin{math} \left|0\right> \end{math} is the neutron 
\begin{math} j=1/2 \end{math} state in one of its two spin projection states 
\begin{math} m_j=\pm 1/2 \end{math} and \begin{math} \left<n\right| \end{math} 
is a delta \begin{math} j=3/2 \end{math} state in one of its 4 spin projection states 
\begin{math} m_j=\pm 1/2 ,\pm 3/2 \end{math}. Since the electric dipole moment 
operator \begin{math} {\bf d} \end{math} is a vector, it can only connect 
\begin{math} j=1/2 \end{math} states to \begin{math} j=3/2 \end{math} states in 
yielding a dipole moment strength  
\begin{equation}
d^2=\left<0\right|{\bf d\cdot d}\left|0\right>=
\sum_n \left|\left< n \right|{\bf d}\left|0\right> \right|^2.
\label{rotn11}
\end{equation}
The neutron polarizability is thereby 
\begin{eqnarray}
\alpha =\frac{2}{3c^2}\left[\frac{d^2}{M_\Delta -M}\right],
\nonumber \\ 
\alpha =\frac{2}{3}\left(\frac{e^2}{\hbar c}\right)\left(\frac{\hbar }{Mc}\right)
\left[\frac{1}{(M_\Delta /M)-1}\right]\Lambda^2 ,
\nonumber \\ 
\Lambda =\frac{2\pi }{3}
\left(\frac{\hbar }{Mc}\right)
\left[\frac{1}{(M_\Delta /M)^2-1}\right],
\nonumber \\ 
\frac{e^2}{\hbar c}\approx 7.29735257\times 10^{-3}.
\label{rotn12}
\end{eqnarray}  
The utility of the quark-diquark string model of fermion bound states, such as the neutron,  
rests on whether the neutron polarizability calculated from this model agrees with experiment. 
The comparison is made below. 

\section{Conclusions \label{conc}}

The theoretical prediction for the neutron polarizability follows numerically from 
Eqs.(\ref{rotn10}) and (\ref{rotn12}),  
\begin{equation}
\alpha({\rm theory})\approx 12.3 \times 10^{-4}\ fm^3.
\label{conc1}
\end{equation}
The experimental value\cite{Schmiedmayer:1991} for the neutron polarizability 
is 
\begin{equation}
\alpha({\rm experiment})=(12.0 \pm 1.5 \pm 2.0)\times 10^{-4}\ fm^3.
\label{conc2}
\end{equation}
We note in passing, that in a similar quark-diquark model of the proton, 
\begin{math} p=duu  \end{math}, the polarizability is expected to have a similar value, 
and it does\cite{Leon:2001}. In that there are no adjustable parameters in the 
theoretical calculation, the satisfactory agreement between theory and experiment 
is unlikely to be fortuitous. Thus, the quark-diquark on a string model pictured in 
FIG. \ref{fig2}, predicting the neutron electric dipole strength in Eqs.(\ref{intro6}) 
and (\ref{intro7}) has a quantitatively adequate significance.

\end{document}